\documentclass{elsart}
\usepackage{amssymb}
\usepackage{graphics}
\begin{document}
\begin{frontmatter}
\begin{center}
\title{Reciprocity theorem in high-temperature superconductors }
\author{Ivan Jane\v{c}ek}$^{1,~2}$,  
\author{Petr Va\v{s}ek\corauthref{cor}}$^{,~2}$
\corauth[cor]{Corresponding author}
\ead{vasek@fzu.cz}
\address
{$^1$University of Ostrava, Dvorakova 7, 70103 Ostrava, Czech Republic\\
$^2$Institute of Physics ASCR, Cukrovarnick\'a 10,\\ 162 53 Praha
6, Czech Republic\\}
\end{center}
\vspace*{10cm}
{\it{corresp. author: P. Vasek, Institute of Physics ASCR, Cukrovarnicka 10,\\ 162 53 Prague 6, Czech Republic\\
tel/fax (+4202) 20318586/ (+4202)33343184\\ e-mail vasek@fzu.cz}}

\newpage

\begin{abstract}
This article is devoted to the problem of the validity of the reciprocity theorem in high-temperature superconductors (HTSC).  The violation of the reciprocity theorem in zero external magnetic fields has been studied.  Experimental data obtained for two different superconducting materials:  BiSrCaCuO and YBaCuO are presented. Results show that the basic form of the reciprocity theorem (without consideration of any additional anisotropy) is not valid  near the critical temperature. We assume that the reciprocity theorem breaking is connected with the existence of an extraordinary transverse electric field originated from additional anisotropy and  more general form of the reciprocity relations should be valid. However, the origin of this anisotropy is not clear yet.   We suggest that the vortex-antivortex dynamics model taking into account  vortex guiding can be responsible for the observed effect. Also the  explanation based on weak P and T symmetry breaking in HTSC which is supported by the observation of the spontaneous magnetisation can not be excluded.
\end{abstract}

\end {frontmatter}

\section{Introduction }
Conventional methods for resistivity measurement use sample with constant cross section and with two planar current contacts on the begin and on the end of the sample. Only potential contacts along sample are point ones.
Multipoint methods use various numbers of  point contacts, which can be used as potential or current ones, respectively. The resistivity tensor can then be derived from resistances $R_{kl,mn}$, which are measured on determined contacts.\\
The most frequent multipoint methods are based on van der Pauw's work \cite{Pauw1}. Original van der Pauw four-point method described in detail in  \cite{Pauw2} can be used for sample of arbitrary shape, and it was generalised also for sample with an anisotropy, which  follows crystallographic symmetry \cite{Pauw3}.  Montgomery method \cite{Montg},  \cite{Log}  uses sample in the form of a rectangular prism with edges in principal crystal directions with contacts placed in the corners to reduce number of the measured resistances, which are necessary for full determination of the resistivity  tensor. Spal \cite{Spal} derived  method for measuring  the resistivity tensor of an anisotropic crystal with magnetic field along crystallographic symmetry axis for prism sample with six point contacts.
These methods differ in their implementation, but the resistances $R_{kl,mn}$ used for determination of the resistivity tensor should obey {\bf the reciprocity theorem}.\\
In the following we will consider four point contacts on circumference of the sample with cyclic notation  A, B, C, and D (see Fig. 1).  Any pair of the contacts can be used as current contacts and remaining pair as potential ones. Thus notation (k, l, m, n) represents any fixed permutation of (A, B, C, D), e.g.  (A, C, B, D).\\
The reciprocity theorem in the absence of magnetic field (and without any other additional anisotropy) declares that the resistances $R_{kl,mn}$ defined as ratio of voltage (measured on potential contacts m and n) and current (put through current contacts k and l) must not change after a current-potential contacts interchange.   This {\bf basic form of the reciprocity theorem} can be written as 
\begin{equation}
R_{kl,mn}= R_{mn,kl}. 
\end{equation}
In the presence of an external magnetic field $\vec{B}$ which introduces an additional anisotropy,  the interchange of the current and potential contacts must be followed by a magnetic field inversion: 
\begin{equation}
R_{kl,mn}(\vec{B})= R_{mn,kl}(-\vec{B}). 
\end{equation}
which is {\bf magnetic field form of the reciprocity theorem}.
This relation is related to the Onsager relation for resistivity tensor. Generally, if  equation of the motion of the observed system is moreover a function of any  additional variable which changes sign after  time inversion ( similar as $\vec{B}$) we must use a generalised form of the reciprocity theorem. In this form we must consider also this additional variable, which must be reverted together with magnetic field. So for samples with internal anisotropy originated from magnetization $\vec{M}$ we can write {\bf generalised form of the reciprocity theorem}: 
\begin{equation}
R_{kl,mn}(\vec{B}, \vec{M})= R_{mn,kl}(-\vec{B},-\vec{M}). 
\end{equation}
M. B\"{u}ttiker \cite{But1}, \cite{But2}, \cite{But3} derived the reciprocity theorem on the basis of microscopic reciprocity of the S-matrix, and demonstrated that this symmetry (Eq. 3) holds for a conductor with an arbitrary number of leads.
Van der Pauw method can be used for the measurement of  resistivity and Hall resistivity of flat samples of arbitrary shape. This is the main advantage and the reason for popularity of this method. Because  of the problem of  preparation of a good "point" contacts on some samples the clover-shaped samples are recommended. In reality the rectangular samples with contacts in the corners are frequently used. The resistivity  can be computed from resistance $R_S$ defined in \cite {Pauw1}.\\
We recall two types of van der Pauw equations 
\begin{equation}
R_{kl,mn} = R_{mn,kl} \hspace{0.5cm} (a), 
\hspace{1cm} -R_{kl,mn} + R_{lm,nk} + R_{km,ln} = 0 \hspace{0.5cm} (b) 
\end{equation}
We define following quantities used for testing the reciprocity theorem in zero magnetic field:
{\bf two-resistance combinations} $R^{\mathrm{(II)}}_{klmn}$  
(for $B = 0$ identical with {\bf deviations from the reciprocity theorem} $D_{klmn}(\vec{B})\equiv[R_{kl,mn}(\vec{B})-R_{mn,kl}(-\vec{B})]/2$)
\begin{equation}
R^{\mathrm{(II)}}_{klmn}\equiv D_{klmn}\equiv[R_{kl,mn}-R_{mn,kl}]/2
\end{equation}
and {\bf three-resistance  combinations}
\begin{equation}
R^{\mathrm{(III)}}_{klmn}\equiv -R_{kl,mn}+R_{lm,nk}-R_{km,ln}
\end{equation}
The Eq. (4a) and Eq. (4b) should be valid in zero magnetic field. However, they may not be valid generally. For magnetic field they must be  modified. Eq. (4a) is the basic form of the reciprocity theorem. Thus in magnetic field it  must be replaced by Eq. (2). 
 Magnetic field equivalent of Eq. (4b) must have on right hand side non-zero values due to transversal field.\\
Now we can go back to situation in zero magnetic field. For example in the presence of the internal magnetisation $\vec{M}$  the general form of the reciprocity theorem (Eq. (3)) should be used with $B = 0$ instead of Eq. (4a).Thus $R_{kl,mn}(0,\vec{M}) \neq R_{mn,kl}(0,\vec{M})$  but general form $R_{kl,mn}(0,\vec{M}) = R_{mn,kl}(0,-\vec{M})$  is valid. Likewise  Eq. (4b) may not be valid and  combination $R^{\mathrm{(III)}}_{kl,mn}$  detects transversal electric field in zero magnetic field - "zero-field effect".   Such effect has been  observed in HTSC as well as in a conventional superconductor. We note that this effect can be induced also by another quantity (a characteristics of the observed system), which changes sign after time inversion (in above equations it formally replaces magnetisation). 

\section{Experiment}
Two types of the high temperature superconductors with preferred c-axis orientation have been prepared. Polycrystalline BiSrCaCuO (2223) with the critical temperature approximately 108 K  was studied. The sample was in the form of a thin disc. The disc was obtained from the pellet of the textured BiSrCaCuO ceramic with preferred c-axis orientation perpendicular to the surface of the pellet.  Also a single domain  YBaCuO (123)  crystal with the critical temperature approximately 90 K  was   prepared by single seed melting method. The sample in the shape of thin square slab was obtained from upper  part of the cylindrical pellet. This part was chosen on the basis of diffraction studies of the pellet to obtain a grain with defined c-axis orientation. The maximal deviation of  axis c from normal to the surface of the sample was less than 5 degrees.\\
We have tested validity of the reciprocity theorem in the form of Eq. (4a) and also validity of the three resistances relation (Eq. (4b)) in both  types of the HTSC. We use above defined quantities $R^{\mathrm{(II)}}_{klmn}$ and $R^{\mathrm{(III)}}_{klmn}$ to demonstrate validity  or invalidity of Eq. (4a) and (4b), respectively.  Data, which are presented in this article, have been obtained in zero external magnetic field. We have performed also experiments in magnetic field, but discussion and treatment of the results of these experiments are more complicated and will be published elsewhere \cite{Jan}.\\
We have measured six resistances (see Fig. 2):
  
(1)	- three resistances in basic cyclic configuration of contacts   (A, B, C, D)\\
	$\hspace*{2cm}R_{AB,CD}, \hspace{1cm}R_{BC,DA} \hspace{0.5cm}\mathrm{and} \hspace{0.5cm}R_{AC,BD}$,\\ \\
(2) - three resistances for cyclic permutation of contacts  (D, A, B, C)\\
	$\hspace*{2cm}R_{DA,BC}, \hspace{1cm}R^*_{AB,CD} \hspace{0.5cm}\mathrm{and} \hspace{0.5cm}R_{DB,AC}=-R_{BD,AC}$.\\
These six resistances allow us  to test the validity of the three-resistance relation (Eq. 4(b)) and of the reciprocity theorem for the cyclic sequence (BC,DA) and the cross sequence (AC,BD) of contact and moreover reproducibility (the first resistance from the first configuration must be identical to the second resistance from the second configuration $R_{AB,CD}=R^*_{AB,CD}$ , or $Z_{ABCD}\equiv R_{AB,CD}-R^*_{AB,CD}=0$.\\
Transport current was commuted to eliminate thermoelectric voltages on the sample and in measuring circuit. We moreover eliminate any other disturbing voltage signal by switching-off the current.


\section{Results of the experiments}
Experimental data obtained on textured polycrystalline BiSrCaCuO (2223) in zero external magnetic field are presented in Fig. 3. Experimental curves for resistances combinations $R^{\mathrm{(III)}}_{ABCD}$,   $R^{\mathrm{(III)}}_{DABC}$ , $R^{\mathrm{(II)}}_{ACBD}$ , $R^{\mathrm{(II)}}_{BCDA}$ ,  and $Z_{ABCD}$   are shown.  Moreover resistance $R_S$ computed from experimental data are plotted as solid line to indicate the superconducting transition. We can see from the behaviour of these curves that the three-resistance relations and the reciprocity theorem are not valid in their basic form near the critical temperature. The $R^{\mathrm{(III)}}_{ABCD}$  and $R^{\mathrm{(II)}}_{ACBD}$  curves are nonzero near to critical temperature with maximum approximately in the middle of the transition of $R_S$. The $R^{\mathrm{(III)}}_{DABC}$  and  $R^{\mathrm{(II)}}_{BCDA}$  are also nonzero near to critical temperature, and moreover, the sign change is observed. Curve $Z_{ABCD}$  is zero for all measured temperature region as expected.\\ 
Experimental data obtained on single domain YBaCuO crystal  is presented in Fig. 4.  Only one three-resistance combination $R^{\mathrm{(III)}}_{AB,CD}$  for various currents through sample is presented. The $R_S$  is also plotted for comparison in this graph. A weak dependence $R_S$ on current is observed in the tail of the transition. Again $R^{\mathrm{(III)}}_{ABCD}$  has nonzero value in the transition region. Moreover, the $R^{\mathrm{(III)}}_{ABCD}$   changes sign from negative to positive for decreasing temperature. We do not observe any pronounced influence of the current on the experimental temperature curve $R^{\mathrm{(III)}}_{ABCD}(T)$ . In Fig. 5 we plot also temperature dependence for individual resistances used in $R^{\mathrm{(III)}}_{ABCD}$  calculation. We note that maximum value of the individual resistances $R_{kl,mn}$ is much greater than difference of the $R^{\mathrm{(III)}}_{ABCD}$  from zero. $R_{AC,BD}(T)$   ("cross combination" of the contacts) changes value from negative to positive near the transition.
Systematic errors due to small shift in the sample temperature may not be negligible for YBaCuO, where the resistance slope in the superconducting transition region is sharp.  A region of the transition is much narrower  (0.5 K) in comparison with BiSrCaCuO (10 K). We have therefore performed a number of tests and other measurement to minimize this effect (see  Fig. 6 and Fig. 7 and discussion latter).\\
 The obtained  data is presented in Fig. 8.  Upper curves show temperature dependence of the both three-resistance combinations $R^{\mathrm{(III)}}_{ABCD}$   and $R^{\mathrm{(III)}}_{DABC}$. Lower curves show equivalent temperature dependence of two-resistance combinations $R^{\mathrm{(II)}}_{ACBD}$  and $R^{\mathrm{(II)}}_{BCDA}$  together with check value $Z_{ABCD}$, which is zero for all temperatures measured.  The basic form of the reciprocity theorem (including the three-resistance formula) is again not valid in the vicinity of the superconducting transition.   

\section{Discussion}
We have observed that the basic form of the reciprocity theorem and the three-resistance formula are not valid in BiSrCaCuO ceramics and single domain YBaCuO crystal.  
The  observed deviations from Eq. (4)  could have following  origins: \begin{itemize}
\item [ 1)]	additional anisotropy due to: 	
   \begin{itemize}
   \item [a)]	spontaneous magnetisation in HTSC near critical temperature
   \item[b)]	guiding  of  created (thermally activated or current induced) vortices and antivortices in preferential direction,
   \end{itemize}
\item [ 2)]non-linearities in VA characteristics
\item [ 3)]	systematic errors of the resistance due to  shift of the  temperature  
     \begin{itemize}
 \item [a)]during temperature stabilisation in one measuring run
 \item [b)]due to different conditions in different runs
 \item [c)]originated from local sample heating at the contacts caused by transport current
     \end{itemize}
\end{itemize}
In the next we will try to discuss all these possible explanations. Let us first estimate the influence of the errors mentioned in 3). Resistance measured on superconductors has typical behaviour with sharp slope near the critical temperature. For a very narrow region of the transition a small shift of the temperature implies large change of the resistance. The transversal electric field  is much smaller than longitudinal one (due to diagonal part of the resistivity tensor) for HTSC. Thus the small error in the longitudinal resistance could influence the determination of  transversal resistances and  the process of testing of the reciprocity theorem.
We suppose that errors, which are presented in point 3a), are not significant in our measurements. We have performed a number of tests, which are focused on study of influence of temperature stabilisation of the sample. We test relaxation processes after temperature settings and also after transport current switching or commuting. We find that at actual setting of measuring device  these processes do not affect measured values.
We did not observe any errors from point  3b)  in BiSrCaCuO in the course of measurement.  We can see good reproducibility ($Z_{ABCD}$ value is zero).  However, differences of the resistances measured on YBaCuO in different temperature cycle deviate from zero near the critical temperature.
The typical situation is presented in Fig. 6, where temperature dependence of deviation of the resistances (from values in the first cycle) is plotted. Near the critical temperature we observe deviations which increase with increasing number of the temperature cycle.
The deviations ( for last cycle) and a derivative of the temperature dependence of the resistances is shown in Fig. 7. We can see good agreement of the behaviour deviations and derivative curves. This means that these deviations can be explained by a small temperature shift ( due to lowering of the level of the cooling medium in cryostat).
From this graph we can deduce that this temperature shift is roughly 0.01 K after about 30 cycles. So for  neighbouring cycles this shift is much less.  This very small temperature shift can be observed in resistance data only because of the sharp slope of the resistance curves near transition.  In final measurement (see Fig. 8) we have eliminated these errors by using resistance data from the same temperature cycle.\\
We have not studied the errors mentioned in point 3c) experimentally in detail, but we suppose that these errors are not significant for following reasons. Temperature of the local heating should increase with an increase of the current through the sample.
Hence, the measured $R^{\mathrm{(III)}}_{klmn}$  data should change value with current change.
We did not observe any current dependence of the $R^{\mathrm{(III)}}_{klmn}$   in YBaCuO (see Fig.4). although the errors by local heating should be more pronounced in YBaCuO due to sharper transition then in BiSrCaCuO.  So, we can consider that the influence of the local heating  is negligible.\\  
BiSrCaCuO superconductor belongs to quasi-2D materials. It is well known that in such matter  non-linear V-A characteristics appears just below $T_C$ due to the spontaneous creation of free vortices and antivortices above  Kosterlitz –Thoules temperature $T_{KT}$, which is  near below $T_C$ \cite{Blatter}. It should be also mentioned that  YBaCuO seems to be rather 3D material where such behaviour is  suppressed. Moreover one  should have in mind that critical currents in BiSrCaCuO are smaller due to weak links and so  current dependencies observed there can be caused by this fact. 
One can  assume that V-A non-linearities could play a role in cause of the reciprocity theorem. However, we do not find any definite information about validity of the reciprocity theorem in system with non-linear dependence of electric intensity on density of the transport current. For example, van der Pauw derivation utilises Ohm's law and Buttiker derivation assumes elastic scattering of the charge carriers only. However, validity of the Ohm's law may not be necessary condition for validity of the reciprocity theorem.
Thus one can suppose that possible influence of the non-linearities should only appear in BiSrCaCuO. For YBaCuO this current dependence does not appear significant. Moreover, in YBaCuO we do not observe in $R^{\mathrm{(III)}}_{klmn}$  any current dependence. Hence, we believe that non-zero values $R^{\mathrm{(III)}}_{klmn}$   and $R^{\mathrm{(II)}}_{ACBD}$ observed  originate from real existence of transversal electric field which appears in the sample even in zero magnetic field. This assumption is supported also by measurement, which have been performed on BiSrCaCuO by means of conventional techniques \cite{Vas2}.\\
We suppose that this additional transversal electric field can be due to additional anisotropy in the sample (not the crystal anisotropy, which has no influence on validity of the reciprocity theorem), related to the  existence of preferential direction defined by vector $\vec{X}$ (for which $\vec{X}(\vec{r}, t) = -\vec{X}(\vec{r}, -t)$).\\
Derivation of the reciprocity theorem (and/or also Onsager's relations) is based on validity of the microscopic reversibility (in time).  The time reversibility is not valid in system with some additional anisotropy, which is due to existence of some preferable direction. For example the direction of the orbital moment in the case of rotating sample, the direction of magnetic induction in the case of the sample in external magnetic field, the direction of the magnetisation of the sample etc. are such  cases. To get symmetry we must revert not only time but also these preferable directions. We suppose that our data can be explained just  this way.\\
We suggest two possible explanations of breaking down the reciprocity theorem (Eq. (4a)) and three-resistance formula (Eq. (4b)) in HTSC. First one is based on existence of induced vortex-antivortex pairs and on application of the guiding vortex model, second explanation is based on observation of spontaneous magnetisation.
As mentioned above we suppose that the reciprocity theorem breaking and violation of three-resistance formula, which is observed in zero magnetic fields, relates to the existence of transversal electric field.  In zero external magnetic fields such transversal field was observed not only in HTSC \cite{Vas2}, \cite{Fra2} but also in some conventional superconductors \cite{Fra1}.   We suppose that this effect - "zero-field effect" is in close relation to the even effect (also improperly called "even Hall effect", observation of the transversal electric field which is an even function of the magnetic induction \cite{Kop} in contradiction to the odd function for true Hall effect ) observed in magnetic field.\\
The theory of the guiding of the vortices was proposed to explain the even effect in superconductors. The theory of the guiding of the vortices \cite{Kop}, \cite{Sta1}, \cite{Sta2}, supposes the existence of a new force  acting on vortex. This force (guiding force) hustles the vortex to move only in given direction. The  guiding force can be explained  as a special case of the pinning force. The direction of vortex motion is determined by the direction of pinning potential valley.  This theory is supported by the fact that an experimental transverse voltage usually conforms to the proposed sin 2$\alpha$ law.    However, the nature of this pinning potential form has not been completely solved.  One of the perspective models is the intrinsic pinning model \cite{Vas1} . This model supposes that the origin of guiding force is due to layered structure of the HTSC. However, other mechanisms (as grain boundary guiding in polycrystalline materials for example) should be kept in mind.\\  
The "zero-field effect" is usually observed near the critical temperature. At first sight it looks rather surprising to observe non-zero transversal electric field without external magnetic field. Near $T_C$ and in magnetic fields such electric field is connected with the existence of free vortices in  the mixed state. But vortices can be generated without application of external magnetic field. In superconductors vortex-antivortex pairs may be excited as thermal fluctuations \cite{Jens}, \cite{Sch}. Or, an isolated vortex and antivortex may be induced by transport current through a sample on the opposite sides of the sample.\\
Glazmann \cite {Glaz} supposes that such induced vortices and antivortices are driven to opposite sides of the sample. The vortices meet oncoming antivortices inside the sample.These vortices and antivortices are attracted and they annihilate in the end. This attraction changes movement of vortex  in direction determined by transport current and antivortex in the opposite direction. Thus both vortex and antivortex motion induce transversal electric field according to the Josephson relation.\\
However, in spite of Glazmann conclusion it seems to us that macroscopic mean transversal electric field must be zero for his model. In zero magnetic fields the system has vortex-antivortex symmetry.  Thus, the probability that vortex on the left meets antivortex on the right-hand side is equal to the probability of the opposite event, from whence mean transverse voltage is zero.  However, a non-zero voltage can be observed, when these induced vortices are moving under influence of a guiding force, for example, the intrinsic pinning force or another. This idea is supported also by validity of the $\mathrm {sin} 2\alpha$ law for the voltage due to zero-field effect \cite{Fra1}, \cite{Fra2}. 
In the next we will analyse models which take into account movement of vortices and antivortices.
The following notation is used: $\vec{v}_L$  and $-\vec{v}_L$ are vortex velocity and antivortex velocity, respectively; $j$ is transport current; $E^V_\perp$ and $E^V_\perp$  are transversal field induced by single vortex and antivortex, respectively, $E_\perp = E^V_\perp + E^A_\perp$  is transversal electric field induced by single splited vortex-antivortex pair; $<E_\perp>$  is mean macroscopic transversal field.

{\it{ Free  vortices and antivortices}} moving with velocity $\vec{v}_L$  and $-\vec{v}_L$  perpendicular to transport 
current cannot induce transversal electric field according to Josephson relation. Thus $E^V_\perp=0$    and $E^A_\perp=0$, hence $E_\perp=0$ and also $<E^V_\perp>$ (see Fig. 9).

{\it{ Glazmann model}} (see Fig.10) considers vortex-antivortex interaction, which can change original direction $\vec{v}_L$  and $-\vec{v}_L$. For simplification we can suppose that single vortex meet antivortex from the left. Probability of this process is $p_L$. The interaction between vortex and antivortex cause that antivortex has a component of the velocity in the direction of the transport current and the vortex in opposite direction.  Thus $E^V_\perp \neq 0$  and  $E^V_\perp=E^A_\perp$, hence  $E_{\perp,L}=2E^V_\perp$. But single vortex can meet antivortex also from the right. Probability of this process is $p_P$.  The interaction between vortex and antivortex cause that now antivortex and vortex   move  in opposite directions than in first case and thus induce opposite transversal field $E_{\perp,P} = -2E^V_\perp$.
Probabilities of  both processes must be equal  $p_L=p_P=1/2$ because system in zero magnetic field has vortex-antivortex symmetry. Thus $<E_\perp>=p_L E_{\perp,L}+p_P E_{\perp,P}=0$.


Our {\it{ guiding force vortex - antivortex model}} (see Fig.11) considers influence of the guiding force on the moving vortex and antivortex. We suppose that angle between the direction $\vec{n}$ of the linear pinning canal and the direction  perpendicular to transport current (original direction of the vortex moving with velocity $v_L$) is  $\alpha$.  Under influence of the guiding force vortex can move with new velocity $\vec{v}'_L= v'_L \vec{n}$, with size $v'_L = v_L \mathrm {cos}\alpha$. Antivortex has velocity of the same size but with opposite direction. Only part of the velocity $(v'_L)_{||}=v'_L\mathrm {cos}(\pi-\alpha)$  parallel with transport current $j$  induce transversal electric field$<E^V_\perp> = (1/2)n_L \Phi_0 v_L \mathrm{sin}2 \alpha$, where $n_L$  is sheet density of the vortices. In this model all vortex move to the left and all antivortex move to the right, hence $<E^V_\perp> = <E^A_\perp>$. In this case the mean macroscopic transversal field is not zero:

\begin{equation}
<E_\perp> = n_L \Phi_0 v_L \mathrm{sin}2 \alpha
\end{equation}

This model can explain existence of the transversal electric field in zero magnetic field and thus also invalidity of the three-resistance formula and maybe also reciprocity theorem. However, on the other hand if the reciprocity theorem (which is a consequence of the Onsager's relation) is not valid, we must suppose existence of any vector $\vec{X}$ defined above. We assume that the internal local field $\vec{B}_I (r, t)$ carried by created vortices and antivortices can play  role of the vector $\vec{X}$ in zero external magnetic field $\vec{B}$.\\
 It should be emphasized that thermal activation of vortex-antivortex pairs should be take into account instead of current induced vortices and antivortices.\\
We can also consider  another explanation of the reciprocity theorem violation. In YBaCuO superconductors, a weak magnetic field ($10^{-5}$ gauss) was detected \cite{Car}, which appears spontaneously at the superconducting transition .  The magnetic signal originated near the edge of  an epitaxial thin film measured.  One interpretation of this observation is that the order parameter carries an intrinsic angular momentum, related to breaking of P and T symmetries. Some experiments on HTSC \cite{Wol}, \cite{Tsu}, \cite{She}  indicate that order parameter has a four-fold $d_{(x^2-y^2)}$  symmetry under rotation of the lattice in contrast to conventional superconductors, whose order parameter is isotropic. Moreover, existence of an imaginary part of the order parameter with s or $d_{xy}$ symmetry  (see \cite{Halp} \cite{Laug} \cite{Salk} and the other references  in \cite{Car}) could break parity (P) symmetry and time (T) reversal symmetry.  The spontaneous magnetisation can be an effect of these breakings.  Furthermore, the reciprocity theorem is a consequence of Onsager's relations, the derivation of which is based on supposition of microscopic reversibility.  In system with additional anisotropy due to magnetisation the basic and/or magnetic field form of the reciprocity is not valid and general formula including the magnetisation reversing must be used. Also authors in \cite{Mar} point out on a consequence of T symmetry breaking: the classic Onsager's reciprocity relation  are not valid which leads to a nonsymmetric conductivity tensor. As a result, the authors in \cite{Halp} mention a possible existence of an analogy to "Hall conductivity" even in absence of an applied magnetic field. In this model the magnetisation $\vec{M}$ can play the role of  additional vector $\vec{X}$.

\section{Conclusion}
We have tested the validity of the reciprocity theorem in HTSC in zero magnetic fields. We find that the basic form of the reciprocity theorem is not valid near the critical temperature. We suppose that this effect results from existence of some additional anisotropy in sample. We suggest two models of the reciprocity theorem breaking and the three resistance formula violating: the vortex dynamics model based on creation vortex-antivortex pairs near the critical temperature and action of guiding  force on them and model based on T-symmetry breaking, which is connected with  the existence of the spontaneous magnetisation near the critical temperature. More general reciprocity theorem should be used even in the case of zero external magnetic field.

\section{Acknowledgements}
This work has been supported by GACR under project No.202/00/1602 and by GAAS under project No. A1010919/99 and by Ministry of education under research plan No. 173100003.\\

\newpage


{\bf Figure caption}

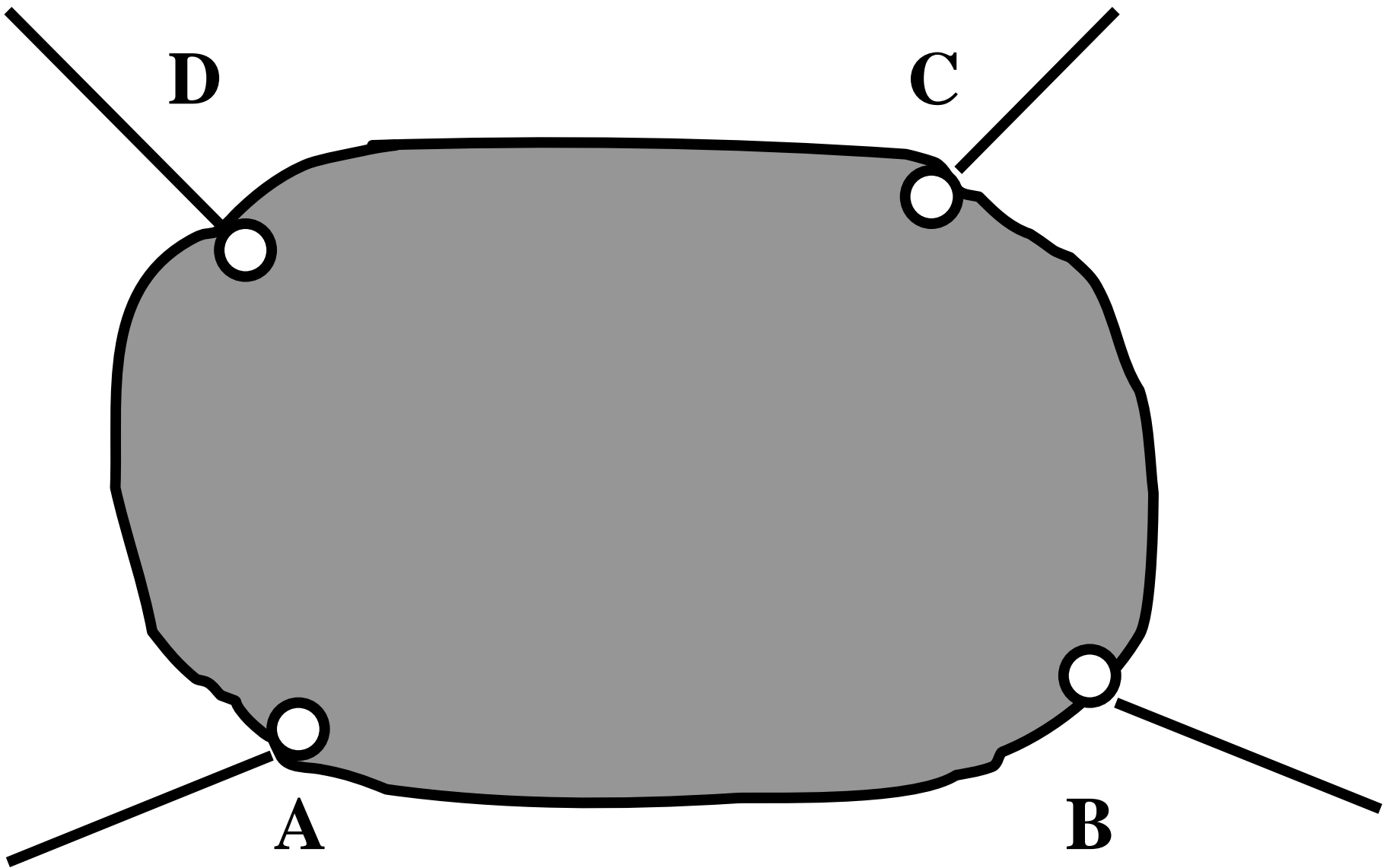
{\bf Fig.1:} {\small Scheme of the sample with four contacts A, B, C, and D placed cyclically on sample 
circumference.}

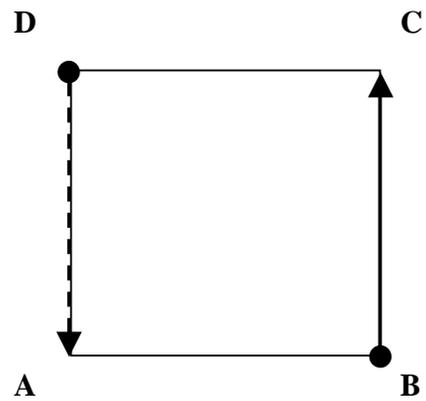
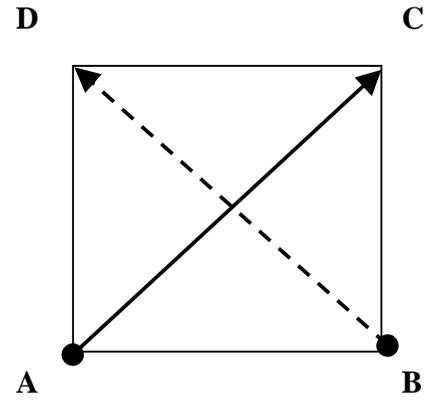
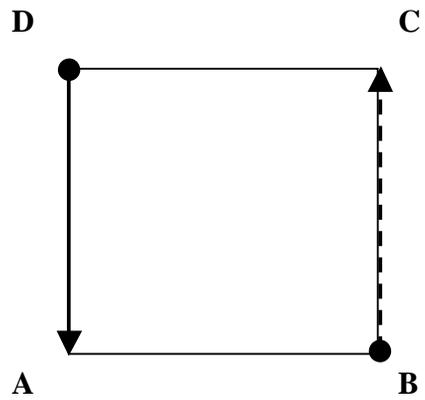
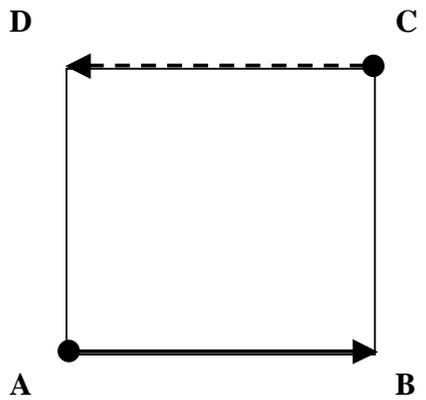
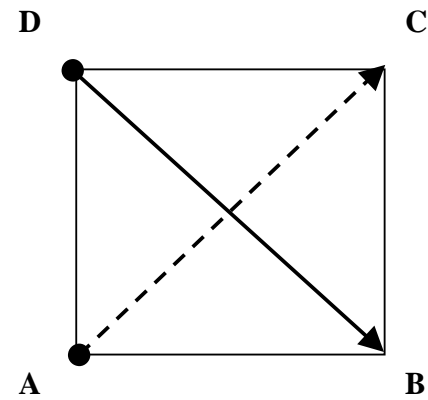
{\bf Fig.2:} {\small Schema of the six  configurations of contacts (kl,mn) used for measurement of the single resistances $R_{kl,mn}$  The solid and dash arrows represent current (kl) and potential (mn) pairs of the contacts, respectively. The bullet denotes minus pole and the arrow end aims at plus pole.  The top three diagrams represent basic configuration. The bottom three diagrams can be obtained by rotation of the top triad.}

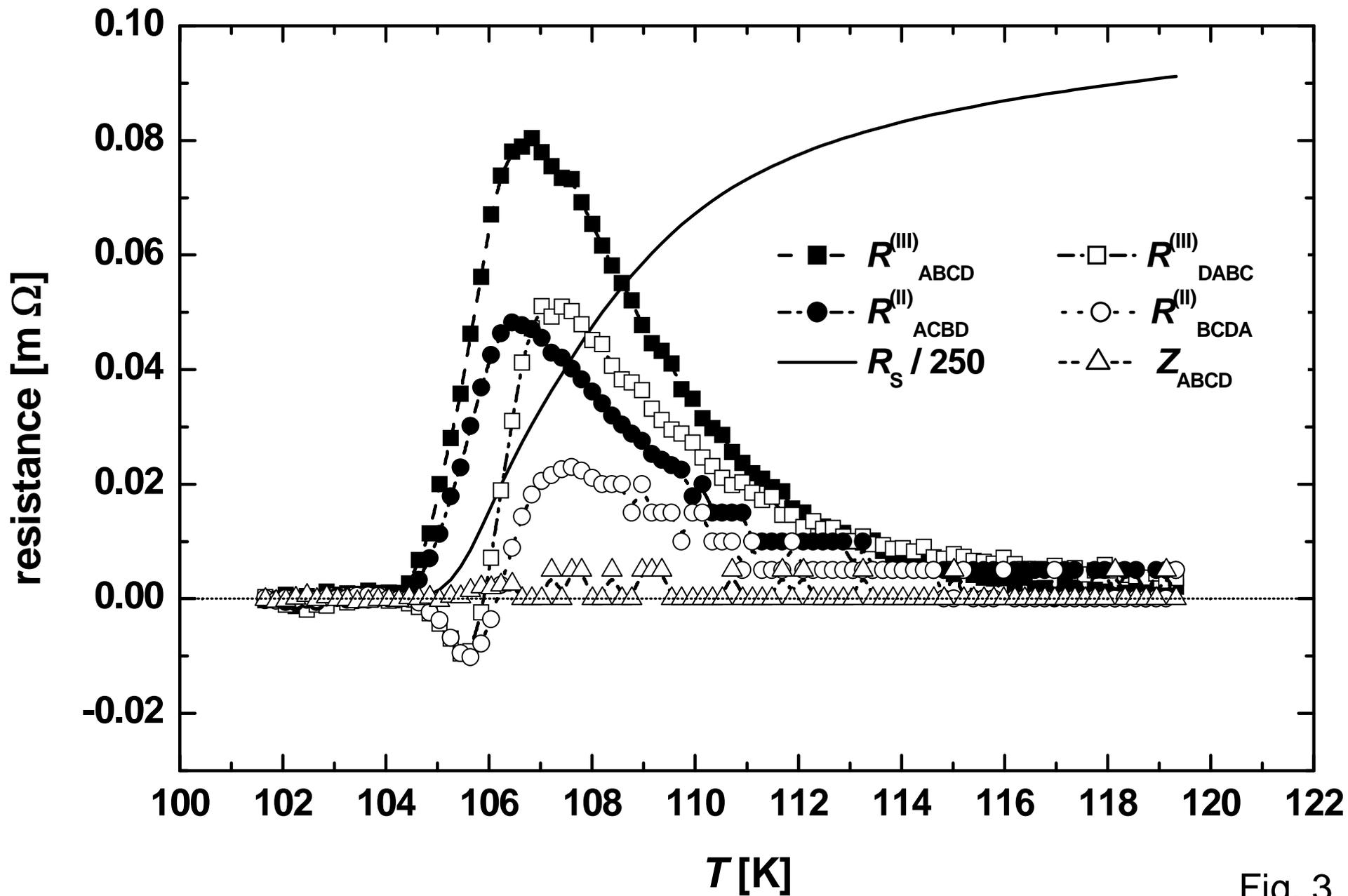
{\bf Fig.3:}   {\small The temperature dependencies of the quantities (defined in the text)  for BiSrCaCuO (2223): the three-resistances combination $R^{\mathrm {(III)}}_{ABCD}$ (solid square) and $R^{\mathrm {(III)}}_{DABC}$ (open square), the two resistances combinations $R^{\mathrm{(II)}}_{ACBD}$ (solid circle) and $R^{\mathrm{(II)}}_{BCDA}$  (open circle), and  $Z_{ABCD}$ (triangle). For comparison also $R_s$ are presented (solid line). Current through the sample was 10 mA.}

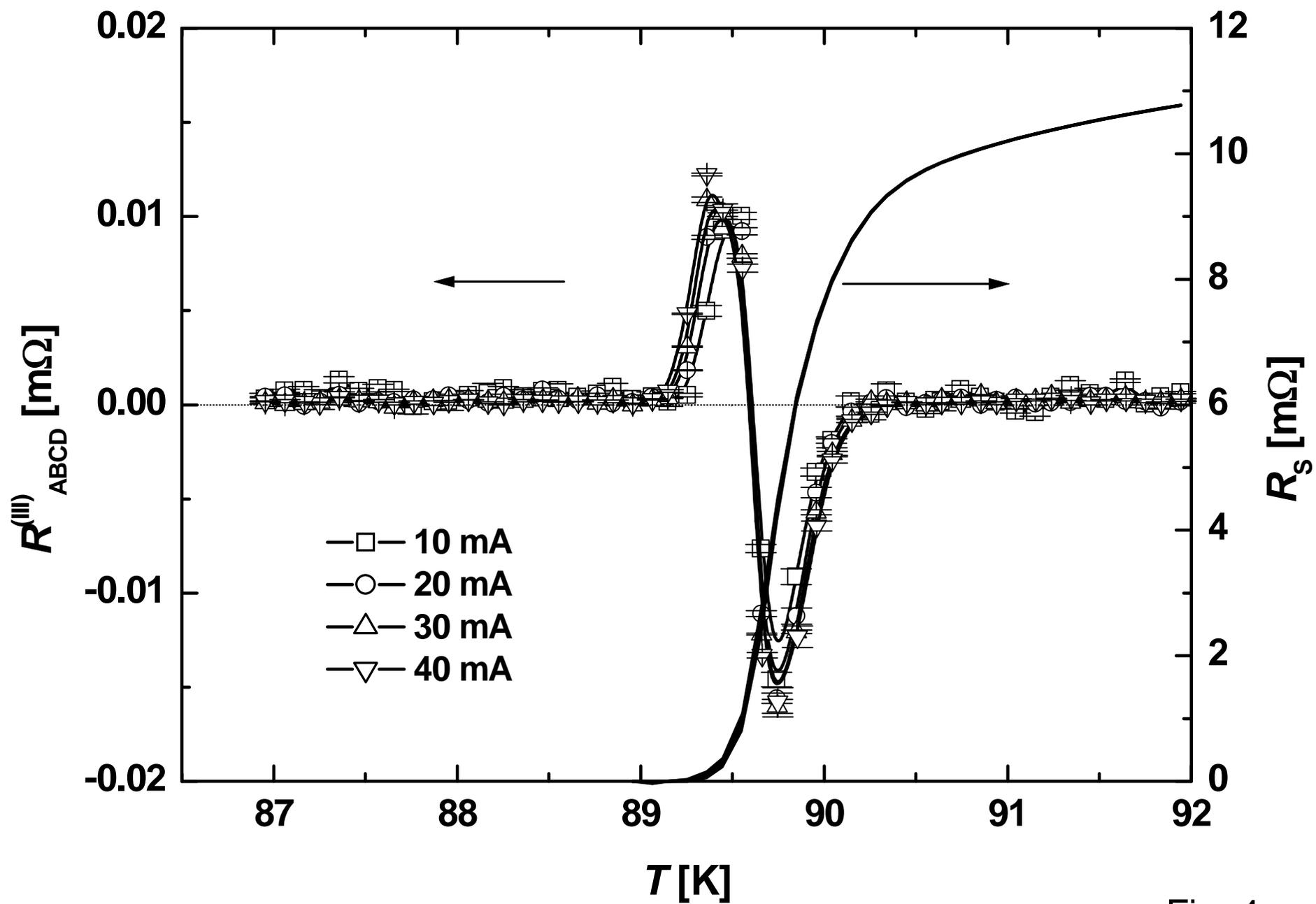
{\bf Fig.4:} {\small The temperature dependencies of $R^{(III)}_{klmn}$ for four values of the transport current for  YBaCuO. For comparison also $R_S$ is shown.} 

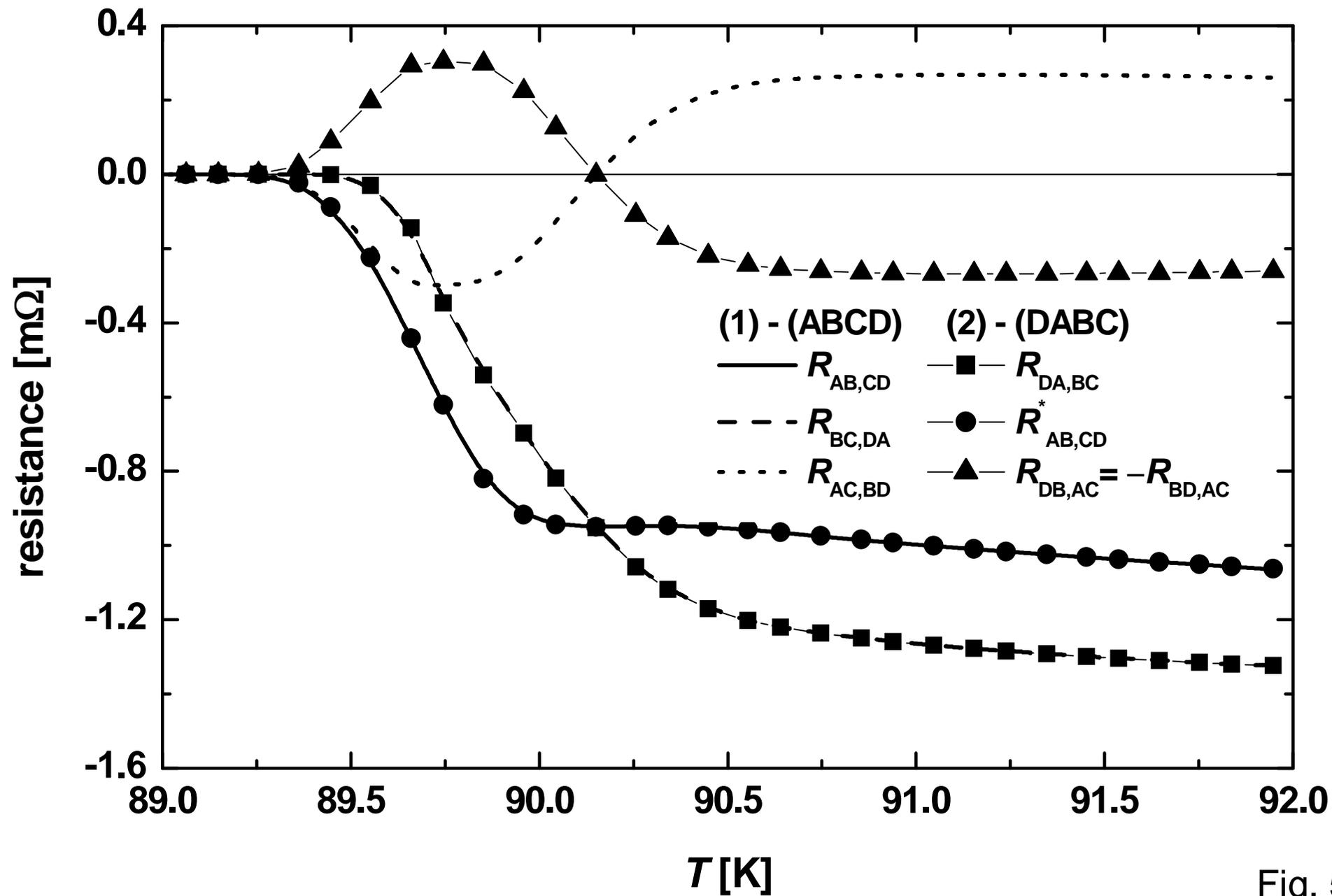
{\bf Fig.5:} {\small The temperature dependencies of the six single resistances $R_{kl,mn}$  for YBaCuO.}

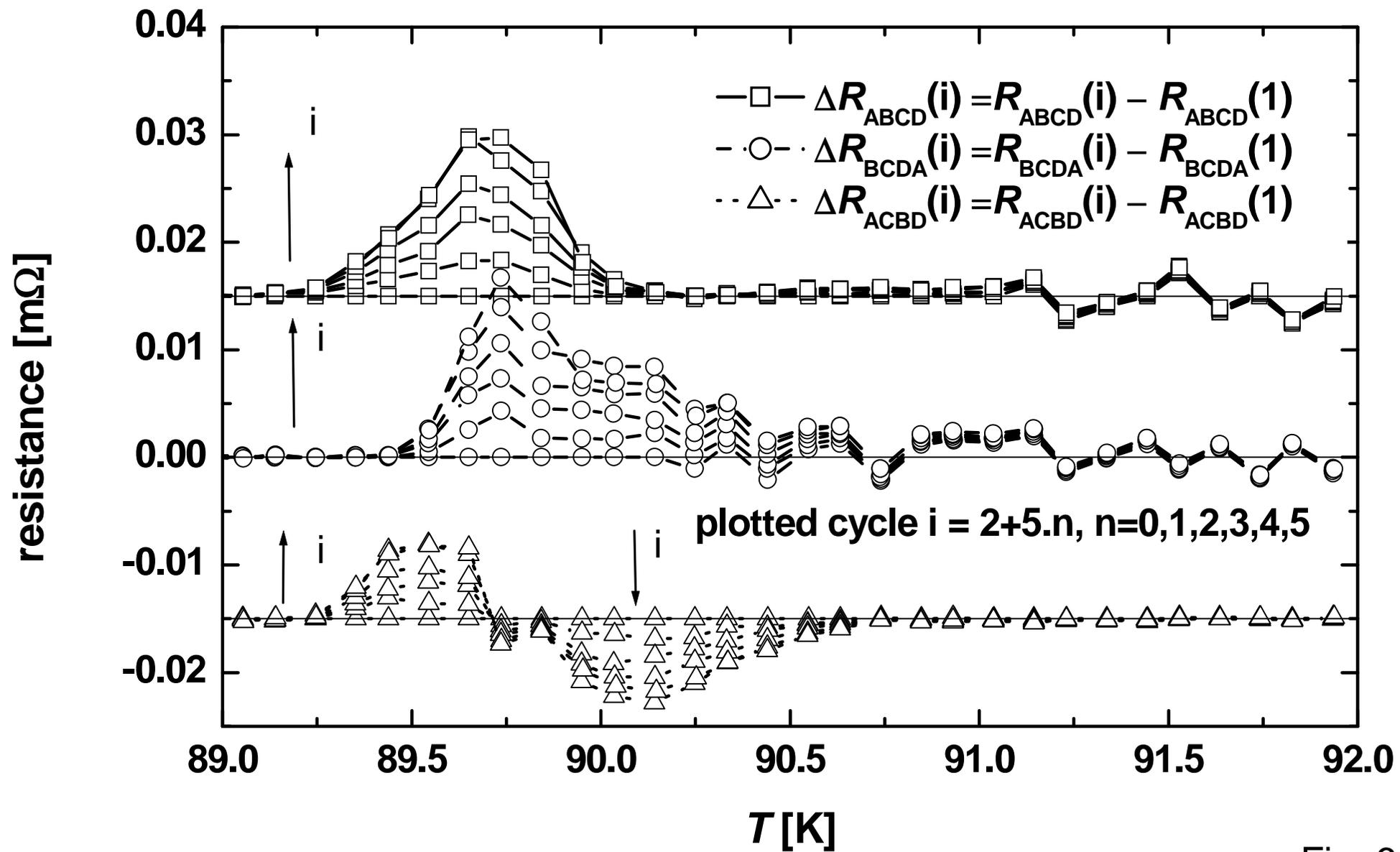
{\bf Fig.6:} {\small Temperature dependencies of the deviations $\Delta R_{kl,mn}$ for subsequent temperature cycles for YBaCuO. The deviation $\Delta R_{kl,mn}$ represents  difference resistances $R_{kl,mn}$  in cycle i  from its values in the first cycle.  For clearity only each fifth cycle is plotted. Current through the sample was 20 mA. Values of the $R^{\mathrm{(III)}}_{ABCD}$  computed from these resistances are presented on Fig. 7}

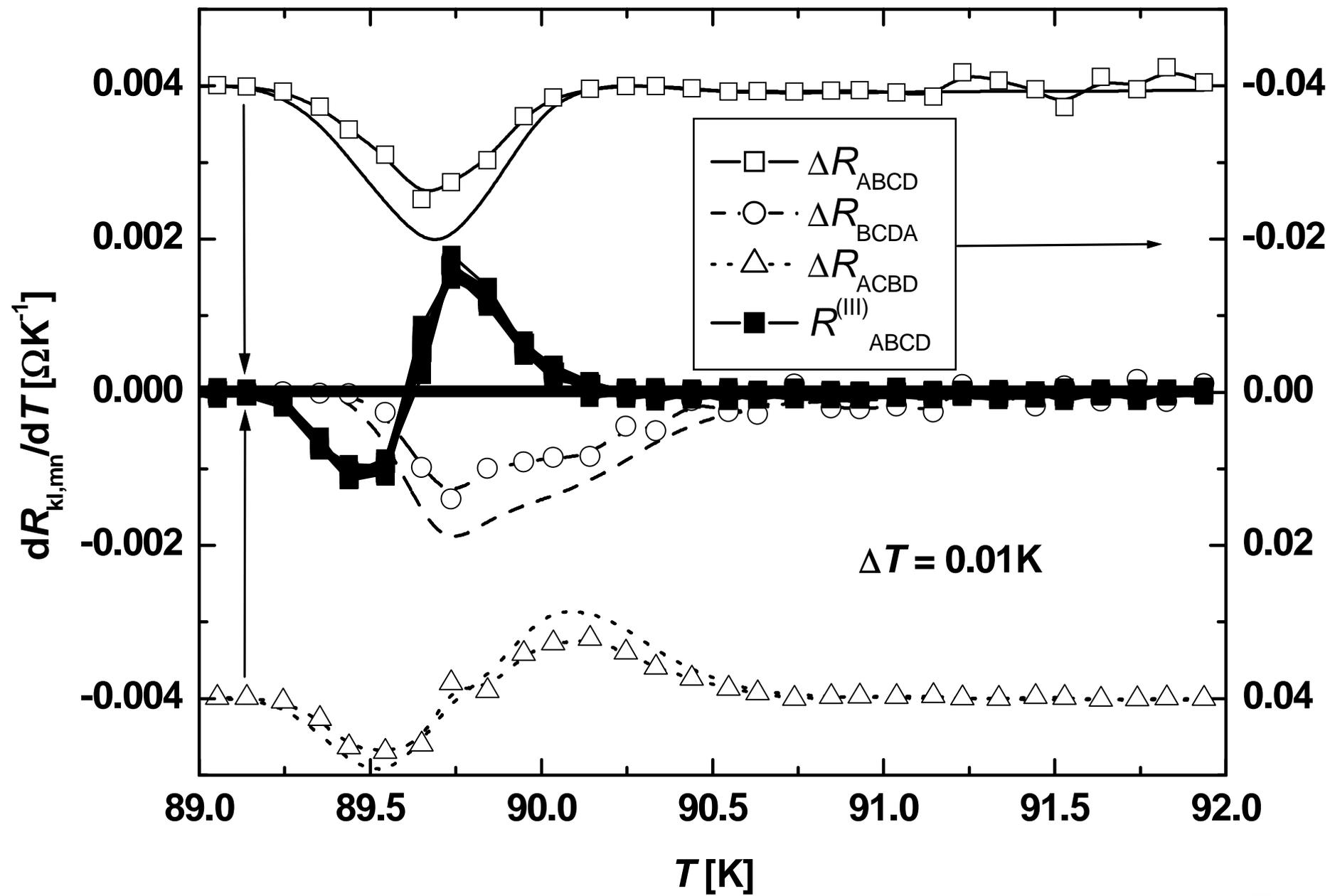
{\bf Fig.7:} {\small Comparison of the deviations $\Delta R_{kl,mn}$ presented in Fig. 6 with derivative $ \mathrm {d} R_{kl,mn} / \mathrm {d} T$ for YBaCuO. Only last cycle deviations (having maximal value) are plotted. Moreover values of the $R^{\mathrm{(III)}}_{ABCD}$  (computed from  the same resistances $R_{kl,mn}$)  are plotted for all cycles. The values of the $R^{\mathrm {(III)}}_{ABCD}$ for different cycle do not differ significantly.}
  
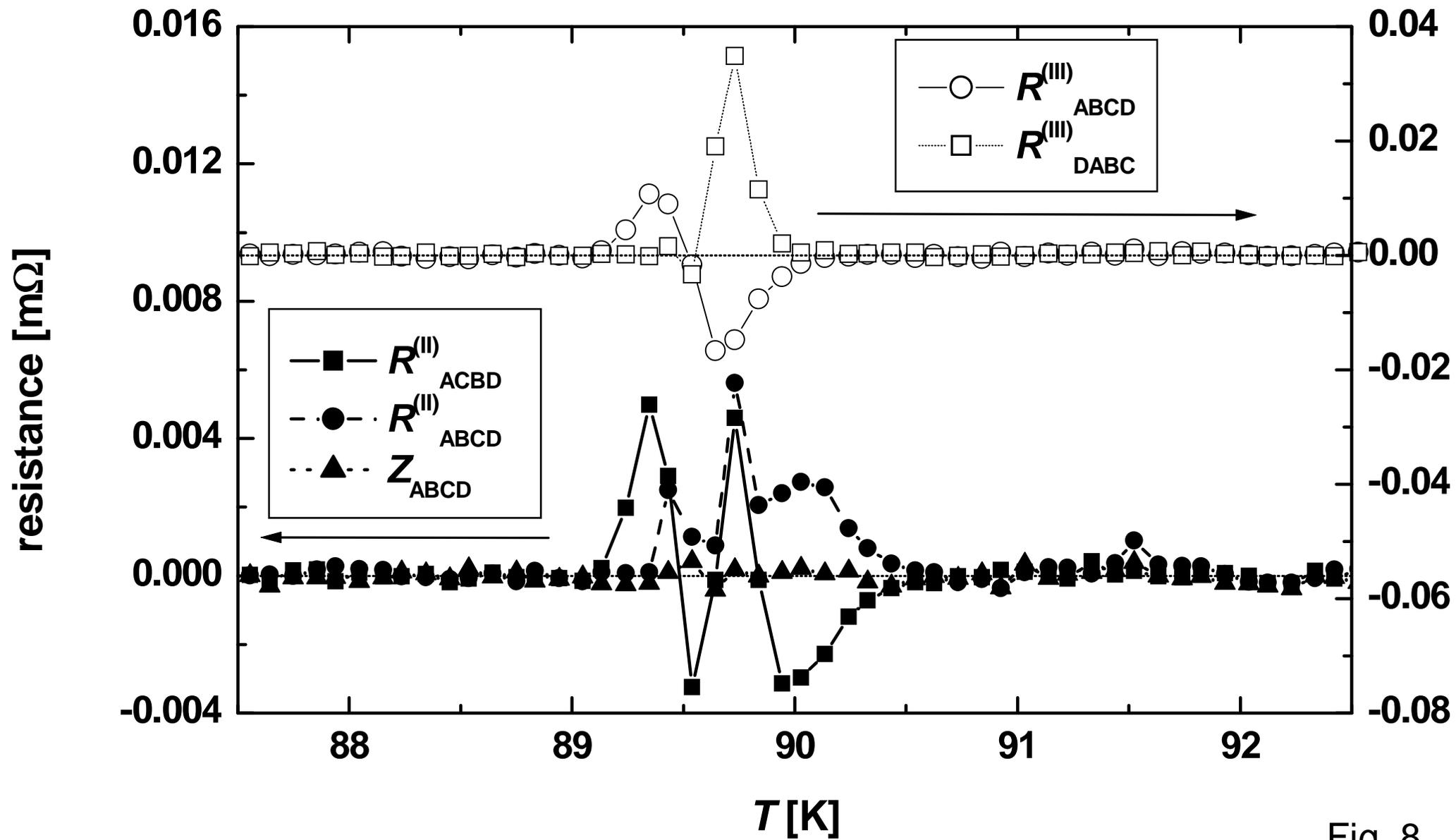
{\bf Fig.8:} {\small The temperature dependencies of $R^{\mathrm {(III)}}_{ABCD}$ and $R^{\mathrm {(III)}}_{DABC}$ (up, open symbols) and $R^{\mathrm {(II)}}_{ACBD}$, $R^{\mathrm{(II)}}_{BCDA}$  and $Z_{ABCD}$ (down, solid symbols). }

{\bf Fig.9:} {\small Free vortices (circles with point) and antivortices (circles with cross) are moving in opposite direction 
perpendicular to the direction (horizontal arrow) of the transport current.}

{\bf Fig.10:} {\small In Glazmann model vortex and antivortex are attracted and change direction of its moving with respect to the  transport current. }

{\bf Fig.11:} {\small Under influence of the guiding force vortex and antivortex are moving along pinning channel. The vortices and antivortices have nonzero component of the velocity longitudinal to the transport current. } 

 
\end {document}
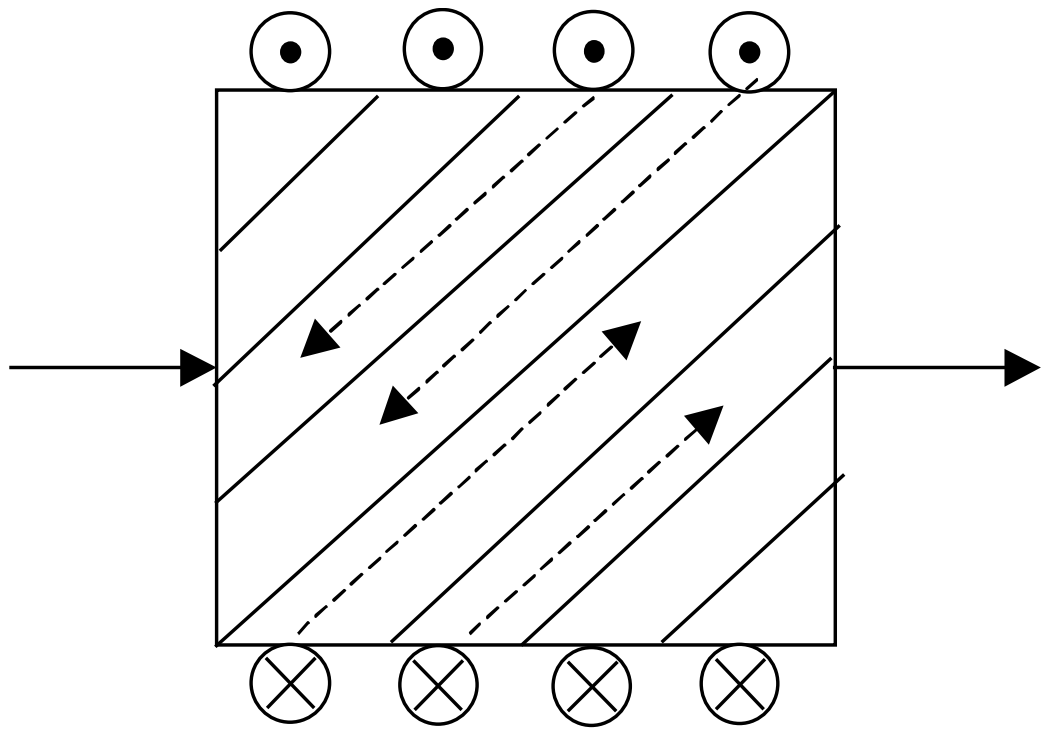
